# Visualization of intervalley coherent phase in PtSe$_2$/HOPG heterojunction

Kai Fan[1], Bohao Li[2], Wen-Xuan Qiu[2], Ting-Fei Guo[1], Jian-Wang Zhou[1], Tao Xie[1], Wen-Hao Zhang[1], Chao-Fei Liu[1], Fengcheng Wu[2,3]* and Ying-Shuang Fu[1,3]†

[1]School of Physics and Wuhan National High Magnetic Field Center, Huazhong University of Science and Technology, Wuhan 430074, China

[2]Schol of Physics and Technology, Wuhan University, Wuhan 430072, China

[3]Wuhan Institute of Quantum Technology, Wuhan 430206, China

†yfu@hust.edu.cn  *wufcheng@whu.edu.cn

**Abstract:**

**Intervalley coherent (IVC) phase in graphene systems arises from the coherent superposition of wave functions of opposite valleys, whose direct microscopic visualization provides pivotal insight into the emergent physics but remains elusive. Here, we successfully visualize the IVC phase in a heterostructure of monolayer PtSe$_2$ on highly oriented pyrolytic graphite. Using spectroscopic imaging scanning tunneling microscopy, we observe a $\sqrt{3} \times \sqrt{3}$ modulation pattern superimposed on the higher-order moiré superlattice of the heterostructure, which correlates with a small gap opening around the Fermi level and displays an anti-phase real-space conductance distribution of the two gap edges. Such modulation pattern and small-gap vanish on the heterostructure of monolayer PtSe$_2$ on bilayer-graphene-covered SiC substrate, due to the increased carrier density in the bilayer graphene. We provide a theoretical mechanism that the $\sqrt{3} \times \sqrt{3}$ modulation pattern originates from the IVC phase of few-layer graphene, which is magnified by the higher-order moiré superlattice. Our work achieves visualization of the IVC phase, and develops an avenue for its generation and amplification via a moiré interface.**

In graphene systems, low-energy electrons possesses both spin and valley degrees of freedom, which, combined with strong electron-electron interactions in certain cases (e.g., in the quantum Hall regime or in magic-angle twisted bilayer graphene) can result in spontaneous emergence of quantum phases accompanied by diverse types of symmetry breaking (*1-11*). Superconductivity has often been observed at the boundaries of these symmetry-broken phases, for example, in Bernal bilayer graphene (BLG) and rhombohedral multilayer graphene (*2, 4, 6*). The origin of superconductivity is a subject of intense interest. One scenario is that the superconductivity may be related to an intervalley coherent (IVC) phase(*12-15*). which has been anticipated as a ground state with spontaneously valley U(1) symmetry breaking within graphene systems (*10, 12, 13, 16, 17*). This particular phase is formed by the coherent superposition of the wave functions of two opposite valleys. In real space, the IVC phase can exhibit a $\sqrt{3} \times \sqrt{3}$ (R3) periodic modulation (*12, 16, 18-20*). The IVC phase, as well as a variety of other symmetry-broken phases, have recently been identified experimentally in graphene systems. However, these observations have mainly been achieved via transport experiments. Direct microscopic visualization of the IVC phase through spectroscopic-imaging scanning tunneling microscopy (SI-STM) is highly desirable to elucidate its physical mechanism, but remains elusive.

In this study, we have successfully visualized the IVC states in a heterostructure composed of highly oriented pyrolytic graphite (HOPG) and monolayer $PtSe_2$ with SI-STM. Monolayer $PtSe_2$ films, that were grown with MBE, form higher-order triangular moiré superlattices on the HOPG substrate. SI-STM measurements resolve a small gap of 140 meV surpassing the Fermi level, and a distinctive R3 pattern with anti-phase real-space pattern around the two gap edges, that are superimposed onto the moiré superlattice. In comparison, monolayer PtSe$_2$ grown on Bernal bilayer graphene covered SiC (BBG-SiC) substrate also display the higher-order triangular moiré superlattice. However, such gap opening and the R3 pattern are absent, but only show a single peak at -190 meV. Our theoretical calculations unveil that the gap opening and the R3 pattern can be a manifestation of the IVC phase in the BBG supported on HOPG that is electrically gated by the interface dipole field from the heterostructure. The interfacial moiré pattern also magnifies the buried IVC phase to a larger lattice period, facilitating its detection. The BBG-SiC has higher electron doping, weaker correlation effects, and thus, cannot form the IVC phase. Our work utilizes a heterorstructure to induce and amplify the elusive IVC phase, opening up a new path way for exploring the correlated physics in graphene systems.

The experiments are performed with a custom-made cryogenic Unisoku STM system. High-quality PtSe$_2$ films were grown by co-depositing high-purity Se and Pt atoms on HOPG and graphene-covered SiC(0001) substrates at 540 K with molecular beam epitaxy (MBE). Detailed descriptions of the experiments are presented in the Supplemental Materials. Figure 1(a) shows a typical morphology of the as-grown PtSe$_2$ islands on HOPG substrate, which contain multilayers of PtSe$_2$. Zoom-in image of the monolayer PtSe$_2$ display a distinct triangular superlattice [Fig. 1(c)], of which orientation and periodicity vary on different PtSe$_2$ monolayer films [Fig. S1]. The superlattice periodicity is predominantly within the range of 0.5 to 0.7 nm, which is larger than the atomic lattice of monolayer PtSe$_2$ (0.370 nm) (*21-23*). The superlattice originates from the moiré pattern formed between PtSe$_2$ and HOPG. To characterize the moiré pattern, we acquired high resolution images to discern the atomic lattice of the HOPG substrate, and the monolayer

PtSe$_2$ coexisting with the superlattice [Figs. 1(b-e)], respectively. Their lattice constants and relative stacking angle can be more clearly seen from the FFT of the atomic resolution image [Figure 1(e)]. This information allows for the simulation of a moiré superlattice, as illustrated in Fig. S2. The simulation shows that the conventional first-order moiré pattern does not align with the observed superlattice. Instead, the reciprocal lattice vectors of the superlattice (**c**) can nicely match the difference between the higher-order reciprocal lattice vector of the PtSe$_2$ lattice (**b**$_{(1,1)}$) and the reciprocal lattice vector of the HOPG lattice (***a***), i.e., **c** = **b**$_{(1,1)}$ − ***a***. This suggests the formation of a higher-order moiré superlattices (*24-26*), as reproduced from the simulation [Fig. S2]. Similar analysis is applied to other sizes of superlattices, which are all well consistent with the higher-order moiré pattern [Figs. S3 and S4]. We note that higher-order moiré pattern has also been observed in some systems (*24-26*), which substantially expands the scope for superlattice formation in heterostructures with large lattice mismatch or large twisting angles.

Having identified the origin of the superlattice, we investigate the electronic properties of the monolayer PtSe$_2$/HOPG heteorstructure. STM image of the monolayer PtSe$_2$ surface at 0.5 V shows the regular triangular superlattice [Fig. 2(a)]. Upon imaging the same field of view at a reduced bias of 0.05 V, the triangular superlattice changes its relative contrast among the otherwise equivalent neighboring sites, as exemplified in three representative neighboring sites colored with red, green and purple spots in Fig. 1(b). This generates a larger $\sqrt{3} \times \sqrt{3}$ period superimposed on the superlattice. The appearance of the larger periodicity suggests the density of states (DOS) of the neighboring sites are different.

Thereafter, we turn to the spectroscopic character of the heterostructure. Tunneling spectrum of the monolayer PtSe2 shows a large insulating gap of ~1.9 eV, whose conduction band minimum (CBM) and valence band maximum (VBM) locate at 0.5 V and -1.4 V, respectively [Fig. 2(c)], conforming to previous studies of monolayer PtSe$_2$ (*23, 27*). Such large-energy scale spectra show no difference on the three representative superlattice sites, implying that the large energy physics is irrelevant to the R3 pattern. As also seen from Fig. S5, a line spectra surpassing multiple superlattice periods indicate the large energy scale spectra are spatially uniform, with no variations in the band gap, and the energies of CBM and CVB.

However, upon zooming into the insulating band gap, the low energy spectra of the three representative superlattice sites all display a small energy gap surpassing the Fermi level, whose gap edges show prominent peaks locating at approximately -0.02 and 0.12 eV, respectively. Spectral intensities of the low energy gap vary among different superlattice sites. Specifically, the spectrum at the purple spot has higher (lower) intensity in its peak at lower (upper) gap edge. In contrast, the spectrum at the red spot exhibits higher peak intensity at its upper gap edge. Such contrast reversal in peak intensity is confirmed from a line spectra [Fig. 2(e)] crossing multiple periods of the superlattice units [white line in Fig. 2(b)]. The line spectra exhibit spatial modulations of the small gap, clearly showing an anti-phase relation between the peak intensities at its upper and lower gap edge. Moreover, such modulation periodicity conforms to that of the R3 pattern.

The above findings can be further substantiated from d*I*/d*V* mappings of the superlattice at different bias voltages. As seen in Fig. 2(f), the d*I*/d*V* mapping at 0.5 V solely shows the triangular superlattice without the R3 modulation, whose unit cell is labelled with red arrows. Its corresponding FFT image has a single set of diffraction points (red circles). In contrast, d$I$/d$V$ mapping at 115 meV and -20 meV, that correspond to energies of the two small-gap edges, both show the R3 modulation (unit cell marked with yellow arrows) superimposed on top of the superlattice [Fig. 2(g, h)]. Evidently, the R3 pattern demonstrates an intensity reversal in its spatial conductance mapping at energies of the upper and lower gap edges. The corresponding FFT displays an additional set of diffraction points (yellow circles) from the R3 pattern, whose intensity gradually weakens with the bias deviating from the gap center [Fig. S6]. These observations suggest that the R3 modulation is only present for low-energy states nearby the gap at the Fermi energy. Consequently, the R3 pattern emerges in STM morphology when imaging at low energies in vicinity to the small gap [Fig. 2(b), Fig. S6], and becomes invisible at high energies deviating from the small gap [Fig. 2(a), Fig. S6]. Such bias dependent evolution of the R3 pattern has also been consistently observed in other studied $PtSe_2$/HOPG superlattices [Fig. S7], confirming the R3 pattern is robust against the variation in the orientation or size of the moiré superlattice.

In view that the small gap and the R3 pattern both emerge inside the large insulating gap of $PtSe_2$, we anticipate those spectroscopic features should be related to the underneath HOPG substrate, where electron correlations can play an important role. The HOPG substrate may be modeled as BBG, to the simplest approximation. Since the correlated states in BBG only occur near the band edge with high DOS and should be sensitive to electron filling, charge doping to the heterostructure is expected to bring critical insight into the possible correlated states therein. For that, we prepare monolayer $PtSe_2$ films on BBG supported on the SiC substrate, whose n-type charge carriers introduce significant electron doping to the atop BBG. Figure 3(a) shows the morphology of the grown PtSe$_2$ islands on BBG/SiC substrate, which resembles to that grown on HOPG. Magnified view of the monolayer PtSe2 also resolves the same triangular moiré superlattices as that on HOPG [Figure 3(b)]. The d$I$/d$V$ spectra of the three neighbouring superlattice sites (red, green and purple dots) consistently exhibit a large insulating gap ~1.9 eV [Fig. 3(c)]. However, the CBM and VBM are shifted rigidly to lower energy by ~300 meV compared to that on HOPG. This amount of energy shift is in reasonable agreement with the energy difference in the Dirac point of HOPG (at ~0 meV) and BBG/SiC (at ~ -300 to -350 meV).

Low energy spectra of the monolayer PtSe2 on BBG/SiC show a single peak at -190 mV without any gap opening, and the spectra are essentially identical on the three typical neighbouring superlattice sites [Fig. 3(d)]. Furthermore, d$I$/d$V$ mapping of the superlattice at the energies of the characteristic peak does not show the R3 pattern either [Fig. 3(e)], where the corresponding FFT [inset in Fig. 3(e)] exhibits only one set of diffraction points. The above spectroscopic features point to distinct contrast to those of monolayer PtSe2 on HOPG. These observations substantiate that electron correlations in graphene systems are responsible for the small gap opening and the R3 modulation pattern.

We now provide a theoretical mechanism for the observed $\sqrt{3} \times \sqrt{3}$ periodic modulation in PtSe$_2$/HOPG heterojunction. In light of the fact that the topography of the HOPG substrate in the experiment is that of a triangular lattice [see Fig. 1(b)], it can be hypothesized that the substrate can be approximated as a BBG. Our first-principles band-structure calculation for PtSe$_2$/ BBG shows that the charge neutrality point of BBG lies within the semiconducting gap of PtSe$_2$, in agreement with the experimental measurement. The PtSe$_2$ layer breaks the spatial inversion symmetry and produces an internal out-of-plane electric field, which opens up a gap at the charge neutrality point in $\pm K$ valleys of BBG. This gap opening suppresses velocities and generates van Hove singularities for band edges, leading to enhanced density of states. States at the valence (conduction) band edges are polarized to the graphene layer adjacent to (away from) PtSe$_2$. Another proximity effect from PtSe$_2$ is that Ising spin orbit coupling are induced for valence band states in BBG. The high density of states in the BBG near the band edge can host strong correlation effects when there are extra charges doped to BBG. Here, we assume that the BBG is hole doped with carriers supplied by the underlying HOPG substrate. With the above setting, we perform self-consistent Hartree-Fock calculation for BBG with an effective applied electric field and hole doping, which indeed lead to IVC states, for example, at a hole density $0.7 \times 10^{12} \text{cm}^{-2}$, as shown by the mean-field band structure in Fig. 4(a).

This IVC state hybridizes the $\pm K$ valleys and results in a $\sqrt{3} \times \sqrt{3}$ density modulation on the graphene atomic lattice, as illustrated by the calculated local density of states (LDOS) in Figs. 4(b)-4(d). Moreover, there is a peak in the density of states right above (below) the Fermi energy, which is associated with the van Hove singularity in the valence band edges. The energy splitting between the two peaks is driven by Coulomb interaction effects and leads to the gap-like feature at the Fermi energy. As illustrated in Figs. 4(c) and 4(d), the peaks above and below the Fermi energy have contrast real space pattern, where sites with the largest LDOS form triangular and honeycomb lattice, respectively. Finally, the higher-order moiré superlattices formed between PtSe$_2$ and BBG magnifies the R3 modulation on the graphene atomic lattice to that on the moiré lattice, which is schematically shown in Fig 4(e). For the comparison system of BBG/SiC, the carrier doping is too high that the Fermi energy is away from the band edge with high density of states, and therefore, strong correlation effects are absent. We consider this theoretical mechanism to be qualitative rather than quantitative. First, we approximate the HOPG substrate as BBG, though similar physics could also arise in other few-layer graphene systems, such as ABA- or ABC-stacked trilayer graphene. Second, the calculated energy splitting between the two peaks above and below the Fermi energy is smaller than the experimentally observed value. This discrepancy may be a quantitative issue influenced by the parameters used in the theoretical model. Additionally, electron tunneling processes through the $PtSe_2$ layer, acting as a barrier, could also affect the determination of the energy splitting. Overall, the proposed mechanism offers a qualitative understanding of the R3 modulation observed in $PtSe_2$/HOPG heterojunctions.

In summary, we have grown monolayer $PtSe_2$ with MBE on HOPG and BBG-SiC substrates respectively, both exhibiting higher-order moiré superlattices. SI-STM measurements unveil that the two heterostructures manifest distinct spectroscopic features. A R3 modulation superimposing over the superlattice is observed on PtSe2/HOPG, which correlates with a low energy gap opening at the Fermi level. Those observations are absent on the $PtSe_2$/BBG-SiC. Our theoretical modeling informed by first-principles band structure illuminate that the atop $PtSe_2$ exerts an electric field to the underneath HOPG. This drives the few-layer graphene system atop HOPG into the IVC phase, exhibiting a R3 modulation in the electron structure. This R3 modulation further manifest itself in the superlattice, as amplified by the interfacial moiré pattern. The IVC phase does not form on $PtSe_2$/BBG-SiC, since its higher doping level deviates the charge neutrality point from the Fermi level. This work not only realizes the microscopic visualization of the elusive IVC phase, but also opens up a new platform for its generation and detection, enriching the capability of moiré interface in the studies of low-dimensional correlated physics.

**Acknowledgement:** We gratefully acknowledge the financial support from the National Key Research and Development Program of China (Grant No. 2022YFA1402400), the National Natural Science Foundation of China (Grants Nos. 92477137, 92265201, U20A6002).

**Note added:** During the preparation of this manuscript, we became aware of two related works posted on arxiv (2411.11163, 2411.14113).

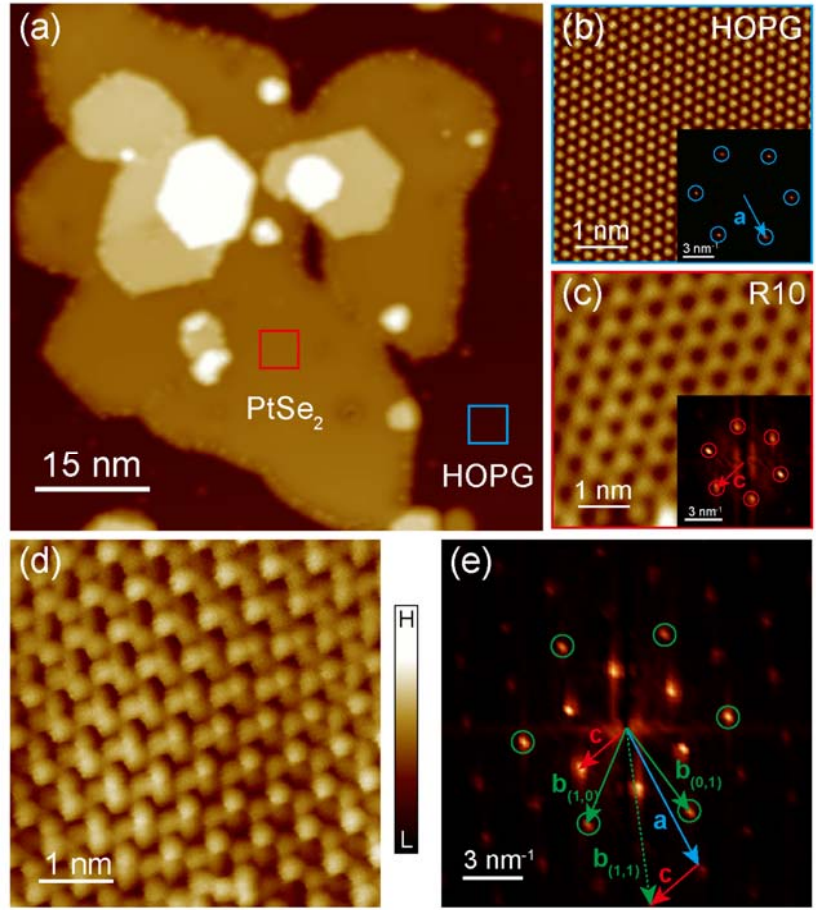

**Fig. 1 Higher-order moiré superlattice in PtSe$_2$/HOPG.** (a) STM topography ($V_b$ = 1 V, $I_t$ = 10 pA) of a PtSe$_2$ island. (b, c) Magnified STM image of the area marked by coloured boxes in (a). (b) is an atomic resolution image of the HOPG ($V_b$ = 0.3 V, $I_t$ = 40 pA), and (c) shows the higher-order moiré superlattice of PtSe$_2$/HOPG ($V_b$ = 0.5 V, $I_t$ = 10 pA). The insets show the corresponding fast Fourier transform (FFT) images. (d) Atomic resolution image ($V_b$ = -0.8 V, $I_t$ = 20 pA) corresponding to (c). (e) FFT image of (d). Red, green and blue solid arrows mark the reciprocal lattice vectors of the higher-order moiré superlattice of PtSe$_2$/HOPG, the atomic lattice of PtSe$_2$ and the atomic lattice of HOPG, respectively. The green dashed arrow marks the higher order reciprocal lattice vector of the atomic lattice of PtSe$_2$.

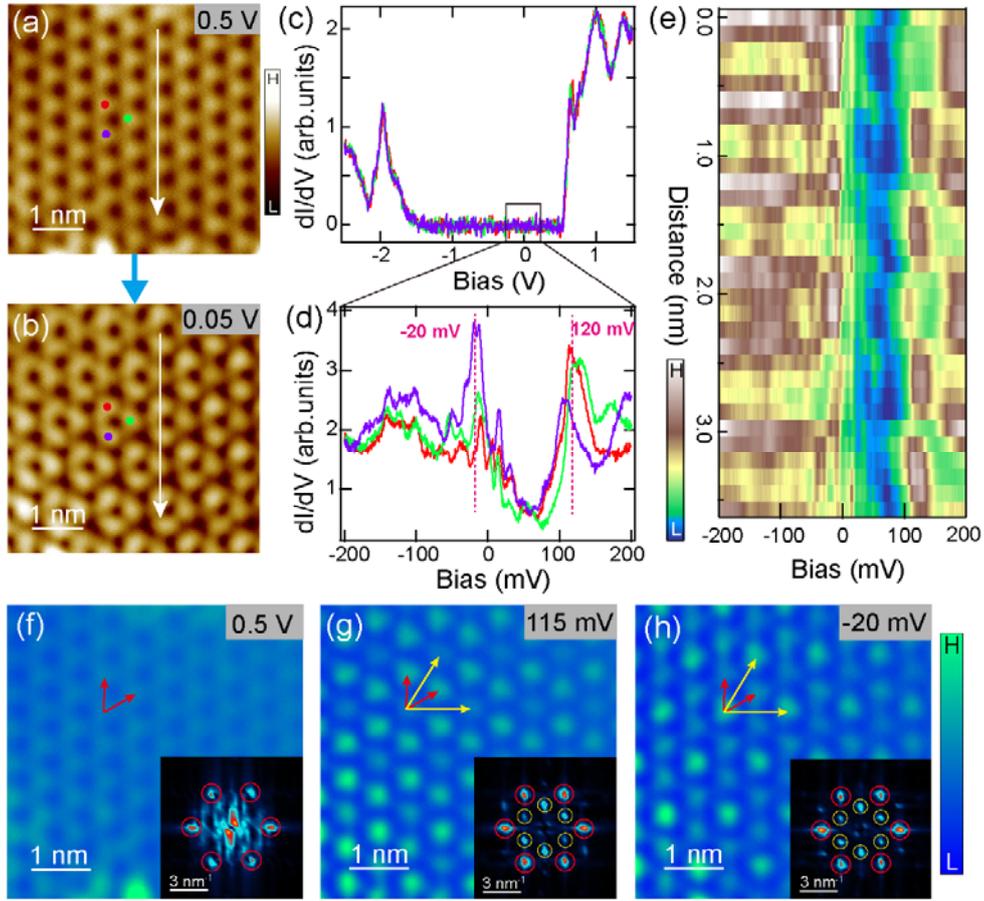

**Fig. 2 $\sqrt{3} \times \sqrt{3}$ reconstruction of the moiré superlattice of PtSe$_2$.** (a, b) STM images of the moiré superlattice measured at 0.5 V (a) and 0.05 V (b), respectively, with the latter exhibiting a $\sqrt{3} \times \sqrt{3}$ modulation. (c) Large-energy range d$I$/d$V$ spectra ($V_b$ = 1.5 V, $I_t$ = 100 pA, $V_{mod}$ = 20 mV) taken at the red, green and purple dots in (a), respectively. (d) Zoomed-in spectra ($V_b$ = 200 mV, $I_t$ = 200 pA, $V_{mod}$ = 2 mV) at the black box marked in (c). (e) Small-energy range d$I$/d$V$ spectra ($V_b$ = 200 mV, $I_t$ = 200 pA, $V_{mod}$ = 2 mV) taken along the white line marked in (a). (f-h) Constant-height d$I$/d$V$ mappings of the same area at different biases. Inset images show the corresponding FFT images. The red arrows mark the moiré superlattice period and the yellow arrows mark the $\sqrt{3} \times \sqrt{3}$ period.

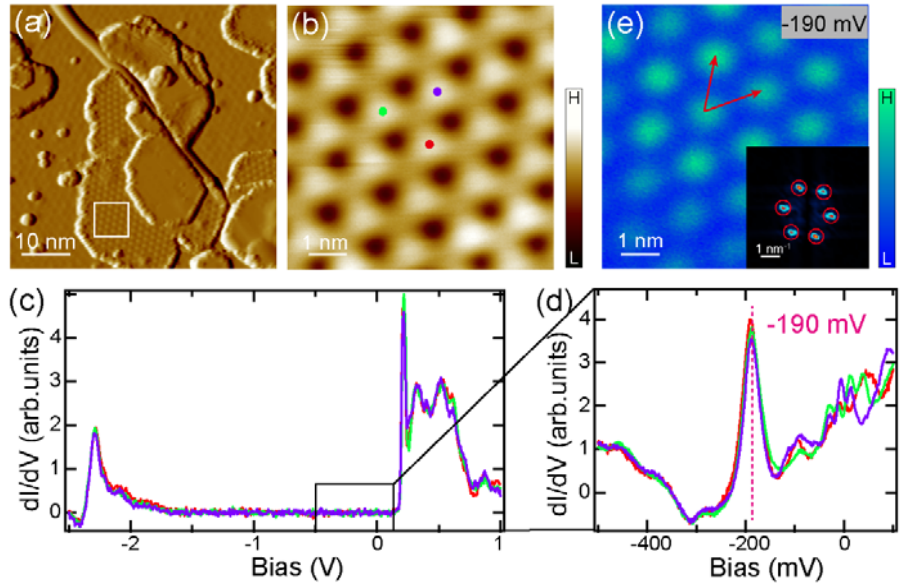

**Fig. 3 Moiré superlattice of PtSe$_2$ deposited on BBG-SiC substrate.** (a) STM topography ($V_b$ = -1 V, $I_t$ = 10 pA) of a PtSe$_2$ island. (b) Magnified STM image ($V_b$ = -1 V, $I_t$ = 10 pA) of the area marked by white boxes in (a). (c) Large-energy range d$I$/d$V$ spectra ($V_b$ = 1 V, $I_t$ = 100 pA, $V_{mod}$ = 20 mV) taken at the red, green and purple dots in (a), respectively. (d) Zoomed-in spectra ($V_b$ = 100 mV, $I_t$ = 100 pA, $V_{mod}$ = 5 mV) at the black box marked in (c). (e) Constant-height d$I$/d$V$ mapping of the same area as (b) at -0.19 V. Inset image shows the corresponding FFT images. The red arrows indicate the moiré superlattice period, but no $\sqrt{3} \times \sqrt{3}$ period are observed.

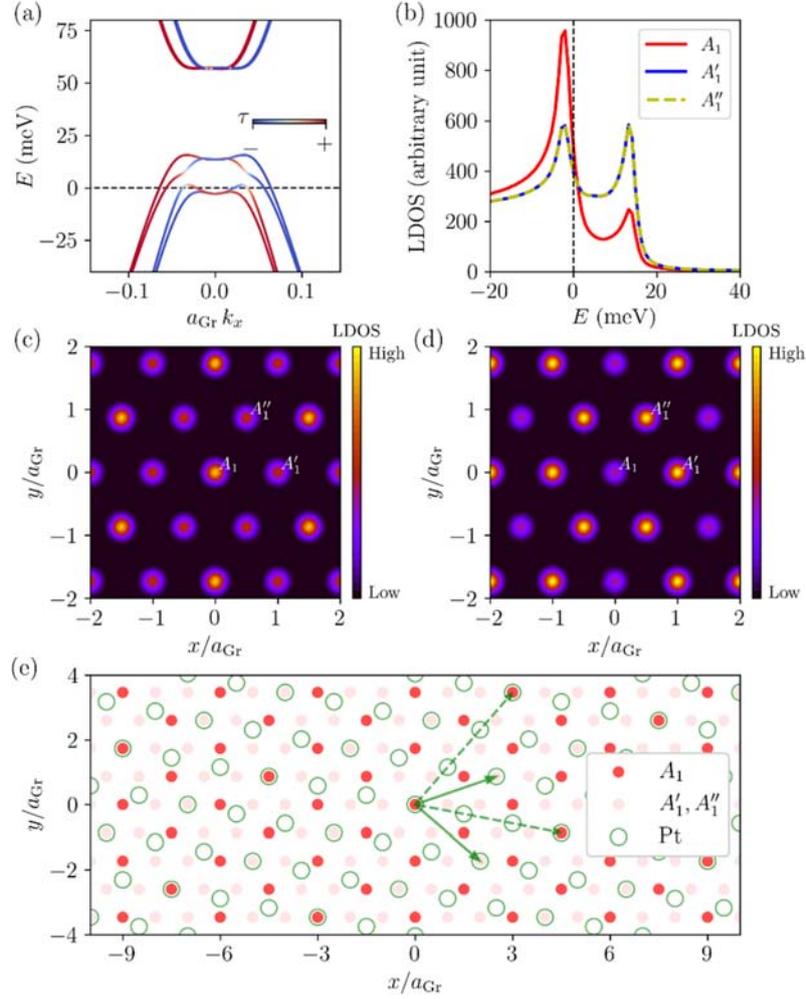

**Fig. 4 Theoretical mechanism for the R3 modulation in PtSe$_2$/HOPG.** (a) Mean-field band structure for the IVC phase in BBG at hole density $0.7 \times 10^{12} \mathrm{cm}^{-2}$. The dashed line represents the Fermi energy, and the colour marks the degree of valley polarization. $a_{\mathrm{Gr}}$ is the graphene lattice constant. (b) LDOS as a function of energy for the band structure in (a) at three carbon sites $A_1$, $A_1'$, and $A_1''$ on the top graphene layer of BBG, which are marked in (c) and (d). The real-space map of LDOS for the peaks below and above the Ferm energy is shown in (c) and (d), respectively, which clearly show the R3 modulation. The $A_1, A_1'$, and $A_1''$ sites would be euqivalent in the absence of the R3 modulation. (e) Illustration of the R3

modulation on the higher-order moiré superlattice induced by that on the graphene atomic lattice. The solid (dashed) arrows mark the primitive lattice vectors of the higher-order moiré superlattice without (with) the R3 modulation.